\documentclass[aps,prl,twocolumn,superscriptaddress]{revtex4}
\usepackage{graphicx}
\usepackage{color}

\def\sbte{Sb$_2$Te$_3$}
\def\bise{Bi$_2$Se$_3$}
\def\bite{Bi$_2$Te$_3$}

\begin{document}
\title{Bulk band structure of Sb$_2$Te$_3$ determined by angle-resolved photoemission spectroscopy}
\author{Henriette E. Lund}
\author{Klara Volckaert}
\author{Paulina Majchrzak}
\author{Alfred J. H. Jones}
\author{Marco Bianchi}
\affiliation{Department of Physics and Astronomy, Interdisciplinary Nanoscience Center (iNANO), Aarhus University, 8000 Aarhus C, Denmark}
\author{Martin Bremholm}
\affiliation{Department of Chemistry,  Interdisciplinary Nanoscience Center (iNANO), Aarhus University, 8000 Aarhus C, Denmark}
\author{Philip Hofmann}
\email{philip@phys.au.dk}
\affiliation{Department of Physics and Astronomy, Interdisciplinary Nanoscience Center (iNANO), Aarhus University, 8000 Aarhus C, Denmark}
\date{\today}
\begin{abstract}
The bulk band structure of the topological insulator \sbte~ is investigated by angle-resolved photoemission spectroscopy. Of particular interest is the dispersion of the uppermost valence band with respect to the topological surface state Dirac point. The valence band maximum has been calculated to be either near the Brillouin zone centre or in a low-symmetry position in the $\bar{\Gamma}-\bar{M}$ azimuthal direction. In order to observe the full energy range of the valence band, the strongly p-doped crystals are counter-doped by surface alkali adsorption. The data show that that the absolute valence band maximum is likely to be found at the bulk $\Gamma$ point and predictions of a low-symmetry position with an energy higher than the surface Dirac point can be ruled out. 
\end{abstract}
\maketitle

\section{Introduction}

\sbte~ was one of the first materials recognised as a three-dimensional topological insulator (TI) with a single Dirac cone surface state \cite{Zhang:2009aa,Hsieh:2009ac}. Its electronic structure is very similar to both that of  \bise~ and \bite, materials that are viewed as prototypical for this family of TIs \cite{Ando:2013aa}. An important point is that both \bise~and \bite~are plagued by degenerate n-doping due to common crystalline defects such that they are ill-suited for experimentally assessing or technologically exploiting transport through the surface states instead of the bulk states. \sbte, on the other hand, is typically strongly p-doped due to the presence of substitutional Sb on Te sites \cite{Horak:1988un}.

As a consequence of the p-doping, only the lower part of the surface state Dirac cone is accessible to angle-resolved photoemission spectroscopy (ARPES) experiments that are routinely used to characterise the predicted surface state dispersion of TIs \cite{Hasan:2010aa} but can only access filled states. This problem can be circumvented by strong surface doping via the adsorption of alkali metals \cite{Seibel:2012aa} or by pumping of electrons into unoccupied states \cite{Reimann:2014aa}. The surface state dispersion can also be characterised by scanning tunnelling spectroscopy in the presence of a strong magnetic field, via the energy spacing of the resulting Landau levels \cite{Jiang:2012aa}. 

The existence of metallic surface states in TIs is enforced by a band inversion, i.e., a swapping of  the $p/d$-like valence band (VB) with the $s$-like conduction band around a high symmetry point, leading to a ``negative'' band gap. More precisely, the relation of the  parity invariants $\delta(\Gamma_i)$ of the eight time-reversal invariant momenta in the bulk Brillouin zone \cite{Fu:2007ac} dictates the $\mathbf{k}$ position of the ``negative'' band gap and the Fermi contour topology of the surface states. For  \sbte~ and most similar crystals, $\delta(\Gamma)$ is -1 while it is 1 for all the other time-reversal invariant momenta, implying a negative band gap at the bulk $\Gamma$ point and that the surface $\bar{\Gamma}$ point is encircled by a single non spin-degenerate surface Fermi contour. 

\begin{figure}
\includegraphics[width=0.45\textwidth]{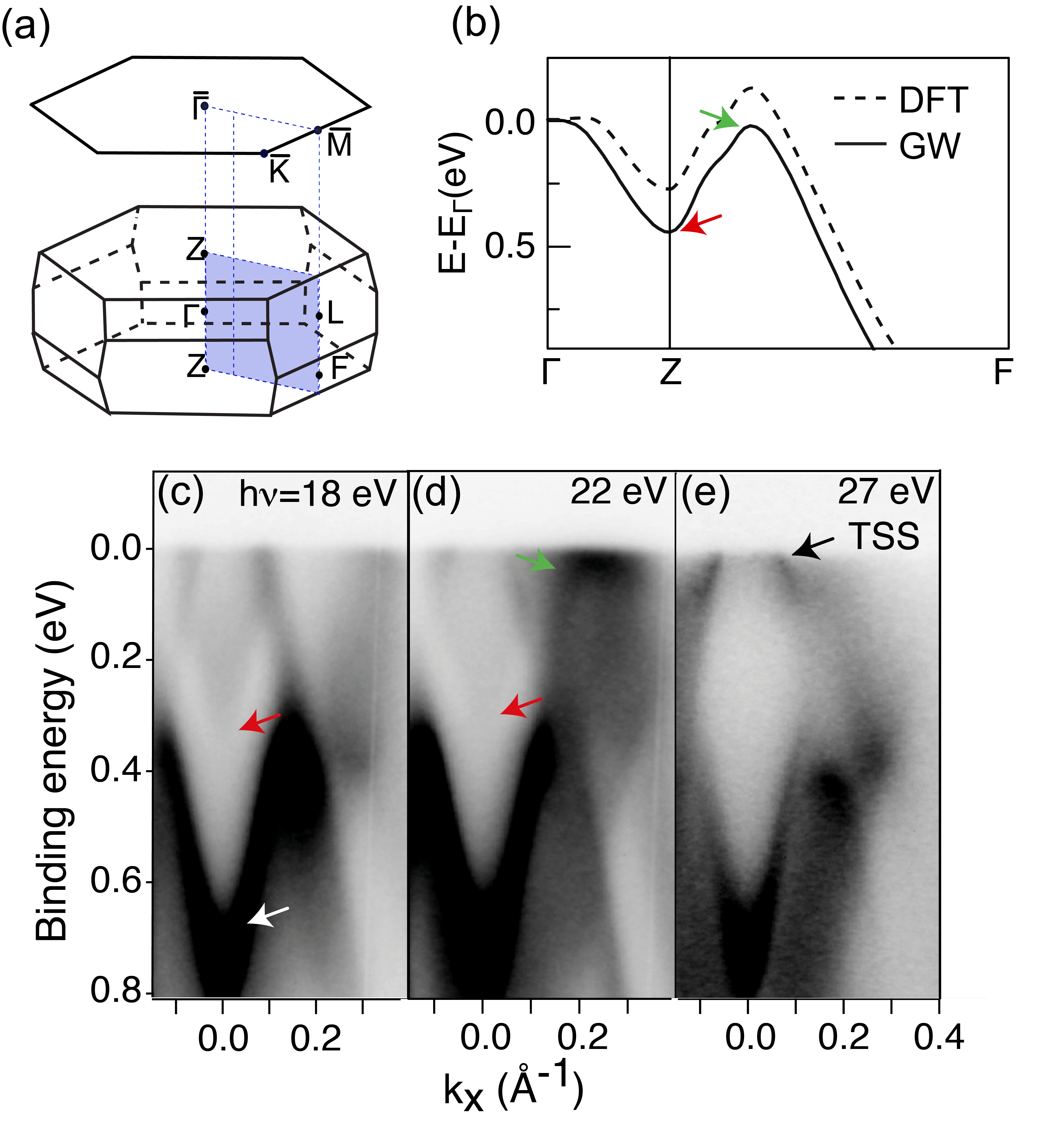}\\
\caption{(a) Bulk Brillouin zone of \sbte~together with the projected surface Brillouin zone. The valence band maximum is predicted to be placed either near $\Gamma$ or in the shaded plane along the $\bar{\Gamma}-\bar{M}$ azimuthal direction. (b) Calculations of the bulk band structure using density functional theory with and without additional GW corrections from Ref.   \cite{Nechaev:2015aa} and \cite{Menshchikova:2011uu}, respectively. The calculations have been shifted so as to align at the band maximum at $\Gamma$, $E_{\Gamma}$, which is also chosen as the energy zero. (c)-(e) Photoemission intensity as a function of binding energy and $k_x$ (in the direction along $\bar{\Gamma}-\bar{M}$), collected at photon energies of 18, 22 and 27~eV, respectively. Several features are marked: the bottom of the valence band along $\Gamma-Z$ (red arrow), the branch of the VB forming a local maximum in the shaded plane (green arrow), the topological surface state (TSS, black arrow) and a further deeper lying surface state (white arrow). The photoemission intensity is scaled such that dark corresponds to high intensity and the colour scale has been saturated to show the relevant features at the Fermi level.}
\label{fig:1}
\end{figure}

The negative band gap has a number of interesting consequences for the effect of many-body interactions on the overall electronic structure of the TIs. In a conventional (non-topological) semiconductor, many-body corrections as implemented in a GW calculation mostly serve to correct the well-known tendency of density functional theory (DFT) to underestimate the size of the band gap. This happens by an approximately rigid shift of VBs and conduction bands away from each other. In a TI, this rigid increase of the band gap still takes place in most of the Brillouin zone (BZ) but not in the region of the negative band gap \cite{Kioupakis:2010aa,Yazyev:2012aa,Nechaev:2013ab,Aguilera:2013aa,Forster:2016ud}. This leads to a $\mathbf{k}$-dependent deformation of the bands which may result in wrong predictions about the band structure and topology \cite{Vidal:2011ul}. \sbte~serves as an excellent example of this.  DFT calculations along high-symmetry directions typically place the absolute  valence band maximum (VBM) between the Z and F points of the BZ  \cite{Zhang:2010ab,Menshchikova:2011uu}, see  Figure \ref{fig:1}(a), (b). The inclusion of many-body corrections such as realised in the GW approximation changes this picture and places the absolute VBM at the bulk $\Gamma$ point \cite{Aguilera:2013aa,Nechaev:2015aa,Forster:2016ud}. Actually, merely displaying the band structure between Z and F may not even show the correct position of the VBM in the DFT calculations because the true VBM could be at any position in the  plane given by the $\bar{\Gamma}-\bar{M}$ direction in the surface BZ  \cite{Nechaev:2015aa} (blue shaded plane in Figure \ref{fig:1}(a)), so that the true VBM in DFT might be even higher than the one between Z and F. 
Note that the band structure differences are not small and can exceed 100~meV, a quarter of the total width of the highest VB along $\Gamma-Z$. A similar interplay between absolute VBM position and many-body corrections takes place in  \bise~\cite{Nechaev:2013aa} and \bite~\cite{Michiardi:2014aa}.

The question of where the absolute VBM is found is not only of interest as a benchmark for many-body corrections in band structure calculations, it also has important consequences for the TIs' transport properties. In particular, the relative position of surface state Dirac point and bulk state dispersion restricts the range of energies in which surface state-dominated transport is achievable \cite{Barreto:2014aa}. In a DFT surface band structure calculation, the absolute VBM in the $\bar{\Gamma}-\bar{M}$ plane lies above the Dirac point of the topological surface state \cite{Hsieh:2009ac,Menshchikova:2011uu}, implying that no level of doping would result in a situation where  electronic transport could be dominated by the surface state electrons near the Dirac point. On the other hand, such surface-dominated transport is possible when the VBM in the $\bar{\Gamma}-\bar{M}$ direction lies below the surface state Dirac point. 

Synchrotron radiation-based ARPES studies were able to map the relevant bulk band structure for \bise~\cite{Nechaev:2013aa} and \bite~\cite{Michiardi:2014aa}, underlining the importance of the corrections in the GW calculations. However, the aforementioned p-doping makes the band structure determination of \sbte~by ARPES more challenging because relevant states can be placed above the Fermi energy in a degenerately doped crystal. Several ARPES investigations from \sbte~single crystal surfaces have been published \cite{Hsieh:2009ac,Seibel:2012aa,Pauly:2012aa,Reimann:2014aa,Seibel:2015wf}, as well as from thin films  \cite{Wang:2010aa,Plucinski:2013vc} where the doping can be controlled by growth conditions \cite{Jiang:2012ab}. The focus of most studies were the topological surface states.  Here we study the dispersion of the topmost VB in order to determine the position and energy of the global VBM. We exploit variable photon energies to explore the complete $\bar{\Gamma}-\bar{M}$ plane. We gain access to the normally unoccupied states by  doping the surface with alkali atoms. A similar approach has been used in Ref. \cite{Seibel:2012aa} in order to access the surface state Dirac point by ARPES. 

\section{Experimental Details}

\sbte~crystals were grown following the procedure outlined in Ref. \cite{Ruckhofer:2021wo}. Attempts were made to synthesize bulk crystals with a smaller p-doping or to achieve chemical bulk counter-doping but these were not successful.  
ARPES experiments were performed on the SGM-3 beamline of ASTRID2 in Aarhus \cite{Hoffmann:2004aa}. Samples were cleaved at room temperature in a pressure better than $5\times10^{-7}$~mbar, prior to the measurement at $2\times10^{-10}$~mbar. The sample temperature during ARPES measurements was $\approx$ 35~K. Energy and angular resolution were better than 50~meV and 0.1$^\circ$, respectively. Rb was dosed from a well-outgassed SAES Getters dispenser during the acquisition of photoemission data. As expected, Rb adsorption leads to a strong electrostatic doping effect but the adsorption on the \sbte~surface is highly unstable. Stopping the Rb evaporation leads to an immediate loss of doping. The Rb coverage was estimated using core level spectroscopy. However, due to the unstable adsorption, this cannot be measured at the same time as the VB spectra but requires a separate cleave of the sample and the evaporation of Rb under the same conditions.  

\section{Results and Discussion}

We start out by presenting data from as-cleaved samples. Due to the p-doping of the material, the VBM at $\Gamma$ and the surface state Dirac point are above the Fermi energy $E_F$ and thus not accessible to ARPES. However, inspecting data from pristine surfaces -- instead of alkali-doped samples -- has the advantage of giving higher quality spectra and the data is thus useful for identifying the salient features and their approximate position in three-dimensional $\mathbf{k}$-space. 
 
Photoemission spectra taken along the $\bar{\Gamma}-\bar{M}$ azimuthal direction (defined as $k_x$) for different photon energies are shown in Figures \ref{fig:1}(c)-(e). 
The highest intensity feature in Figures \ref{fig:1}(c)-(e) is a surface state in a projected bulk band gap of the VB (marked by a white arrow). This state and its Rashba-splitting have been studied in detail in Refs. \cite{Pauly:2012aa,Seibel:2015wf}.  A second surface state is the metallic topological state forming a Dirac cone around the $\bar{\Gamma}$ point (marked as TSS). Due to the p-doping, only the lower part of the Dirac cone is visible, most clearly in Figure \ref{fig:1}(e). 

The bulk features of interest are the highest VB at $\bar{\Gamma}$ and a possible further VBM in the  $\bar{\Gamma}-\bar{M}$ plane. Starting with the former,  a GW calculation of the highest VB gives an absolute energy difference of only $\approx$450~meV between the $\Gamma$ and $Z$ point of the BZ \cite{Nechaev:2015aa} with the highest binding energy reached at the $Z$ point (see Figure \ref{fig:1}(b)). This total band width can be small compared to the energy broadening due to the short inelastic mean free path of the photoelectrons and the resulting uncertainty in  $k_z$, so that the entire VB can be visible as a broad feature, as previously observed for this material \cite{Pauly:2012aa}. Nevertheless, the data taken using a photon energy of  $h\nu=$18~eV actually shows a weak band reaching a binding energy of $\approx$300~meV between the branches of the topological surface state (see red arrow in Figure \ref{fig:1}(c)). We assign this to the bottom of the VB. The data taken at 22~eV also shows this band, or at least a continuum with an edge following the same dispersion (red arrow in Figure \ref{fig:1}(d)). Given the fact that the upper VB reaches its highest binding energy at these photon energies, it is possible to assign the spectra taken at $h\nu=$18~eV and $h\nu=$22~eV to a region around the $Z$ point of the BZ. The detailed $k_z$ dispersion of the features will be clarified below. The spectrum in Figure \ref{fig:1}(e) taken at 27~eV photon energy is entirely different. The topological surface state is much more pronounced and there is little photoemission intensity between the surface state branches. The qualitative similarity to data taken from \bise~(see online supplementary material of Ref. \cite{Bianchi:2010ab}) suggests that the data is taken near the $\Gamma$ point of the BZ where the top of the VB should be visible. However, in \bise, the VBM at $\Gamma$ is never actually observed because its dispersion almost coincides with the lower part of the topological surface state Dirac cone. 
   
The highest VB in the  $\bar{\Gamma}-\bar{M}$ plane is seen in Figure \ref{fig:1}(c) as a diffuse intensity at $E_F$ and in Figure \ref{fig:1}(d) as a band moving up to $E_F$ (marked by a green arrow). The band does not appear to cross $E_F$ but rather levels off and shows a flat dispersion $\approx 30$~meV below $E_F$. Finding this band at the photon energies corresponding to $k_z \approx Z$ for normal emission (rather than to $k_z \approx \Gamma$) is expected from its calculated position in the bulk BZ (see dispersion along $Z-F$ in Figure \ref{fig:1}(b)) \cite{Nechaev:2015aa}.  When mapping the photoemission intensity at the Fermi level in two-dimensional $\mathbf{k}$ space, this band gives rise to six ``flower petals'' along $\bar{\Gamma}-\bar{M}$ (this can be seen in Figure 2(a) in Ref. \cite{Ruckhofer:2021wo} using data taken from the same crystal as here at $h\nu=22$~eV, and in Refs. \cite{Seibel:2012aa,Plucinski:2013vc}). 

\begin{figure}
\includegraphics[width=0.50\textwidth]{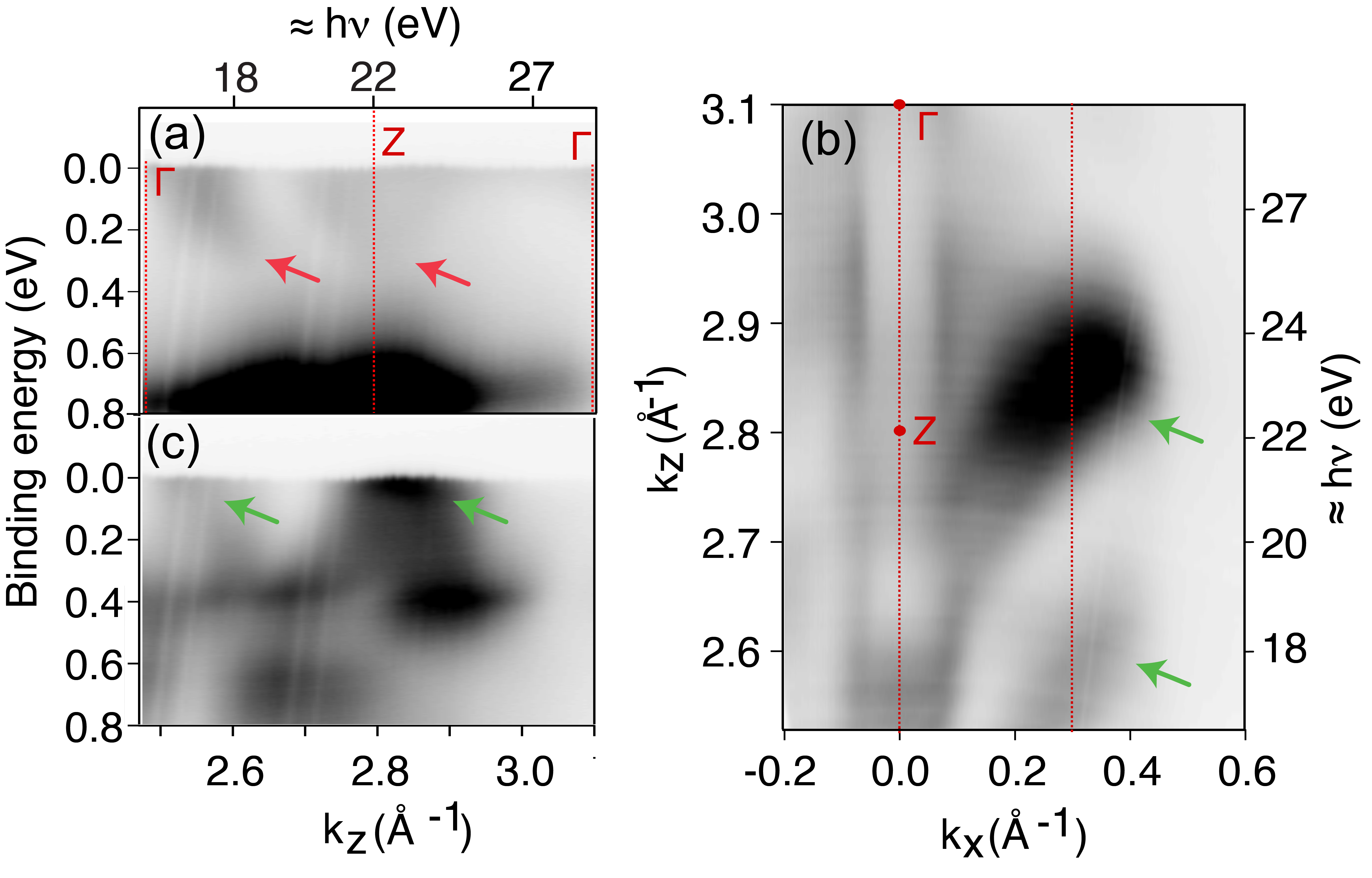}\\
\caption{(a) Photoemission intensity as a function of binding energy and $k_z$ in normal emission. The spectrum is integrated along $k_x$ within $\pm$ 0.02~\AA$^{-1}$ around normal emission. (b) Photoemission intensity at the Fermi energy as a function of $k_z$ and $k_x$. The spectrum is integrated within $\pm$ 100 meV around the Fermi level. In panels (a) and (b), the $h\nu$ scale is strictly correct for normal emission and the Fermi energy and otherwise only  approximate. (c) Photoemission intensity as a function of binding energy and $k_z$ for $k_x$ = 0.3 \AA$^{-1}$, integrated within $\pm$ 0.02~\AA$^{-1}$ around this value. The colour scale is chosen such that some features are saturated.}
\label{fig:2}
\end{figure}

A more systematic investigation of the $k_z$-dependent bulk band structure in the $\bar{\Gamma}-\bar{M}$ plane is shown in Figure \ref{fig:2}. Here, $k_z$ of the outgoing photoelectrons and thereby also the $k_z$-dependence of the initial states is explored by collecting photon energy-dependent data. The  photon energies have been converted into $k_z$ values using a free electron final state model. 
Assuming a work function of $\approx 5.0$~eV \cite{Hao:2012wt} and an inner potential of $V_0=12.8$~eV reproduces the position of a $Z$ point at $h\nu=22$~eV. The corresponding $k_z$ value is 2.795~\AA$^{-1}$ which is 9 times the bulk $\Gamma-Z$ distance. The same $Z$ point is observed at a photon energy of $\approx 26.6$~eV for \bise. For higher photon energies, we find  indications of $Z$ points at $h\nu=54.4$ and 74.8~eV (not shown). These also agree with the free electron final state model. 

Figure \ref{fig:2}(a) shows the photoemission intensity as a function of binding energy and $k_z$ collected in normal emission. The data are similar to the corresponding observation for \bise~(see Figure 6 in Ref. \cite{Bianchi:2012ac}). A broad parabolic feature is visible near $E_F$, reaching its highest binding energy at $k_z \approx$ 2.8~\AA$^{-1}$ (red arrow). We identify this as the bottom of the uppermost VB at $Z$. A similar structure is visible at lower $k_z$, reaching the highest binding energy at around 2.6~\AA$^{-1}$ (also marked by a red arrow). These observations are consistent with the position of the uppermost VB in Figure \ref{fig:1}(c) and (d) since the two $k_z$ values correspond to photon energies of 18 and 22~eV, respectively. However, the free electron final state model cannot explain the observation of the upper VB minimum at $h\nu=18$~eV ($\approx$ 2.6~\AA$^{-1}$, as seen in Figures \ref{fig:1}(c) and \ref{fig:2}(b)), as this $k_z$ value does not correspond to a $Z$ point. A similar feature for \bise~was assigned to a surface umklapp process \cite{Bianchi:2012ac}. The same explanation may hold here but the emission could also occur via the non-free electron like nature of the final states at the low photon energies used here.  

Having identified the bulk $Z$ point in the data, we move our attention to the VBM in the $\bar{\Gamma}-\bar{M}$ plane that is expected to be found at a similar $k_z$. To this end, Figure \ref{fig:2}(b) shows the photoemission intensity at the Fermi energy as a function of $k_z$ and $k_x$ along $\bar{\Gamma}-\bar{M}$. Pronounced features are the two vertical streaks that correspond to the Fermi level crossings of the topological surface state. In addition, two dispersive broad structures are visible around $k_x =0.3$~\AA$^{-1}$ and at $k_z$ values around 2.6 and 2.8~~\AA$^{-1}$ (marked by green arrows). These stem from the VBM in the $\bar{\Gamma}-\bar{M}$ plane and are also seen in Figures \ref{fig:1}(c) and (d). The shape of the high intensity features agrees remarkably well with the calculated constant energy contours for the VB in Figure 3 of Ref. \cite{Nechaev:2015aa}. Figure \ref{fig:2}(c) shows the dispersion of these structures (also marked by green arrows) with $k_z$ by plotting the photoemission intensity as a function of photon energy in the same way as in Figure \ref{fig:2}(a) but for $k_x =0.3$~\AA$^{-1}$. The dispersion with $k_z$ is evident. As in the selected data of Figure \ref{fig:1}(d), it appears as if the band barely reaches the Fermi level without actually crossing it. This view is also supported by the lack of sharp structures in Figure \ref{fig:2}(b) such that the ``islands'' of high intensity can be interpreted as a spill-out of the intensity from band maxima below $E_F$, rather than distinct crossings.

In order to confirm this interpretation and to investigate the position of the VB along $\bar{\Gamma}-\bar{M}$ further, we gain access to the usually unoccupied band structure by adsorbing a low coverage of Rb atoms onto the surface while collecting ARPES data. This leads to doping with electrons and to a downward band bending at the surface \cite{Bianchi:2012ab}. We can track the occupation of the topological surface state and the normally unoccupied VB, and even populate the bottom of the conduction band. We perform the doping experiment at two photon energies, 22~eV and 18~eV to be close to the  $Z$ point expected for free electron final states at 22~eV and to the $Z$ point-like features reached at 18~eV. 

\begin{figure}
    \includegraphics[width=0.45\textwidth]{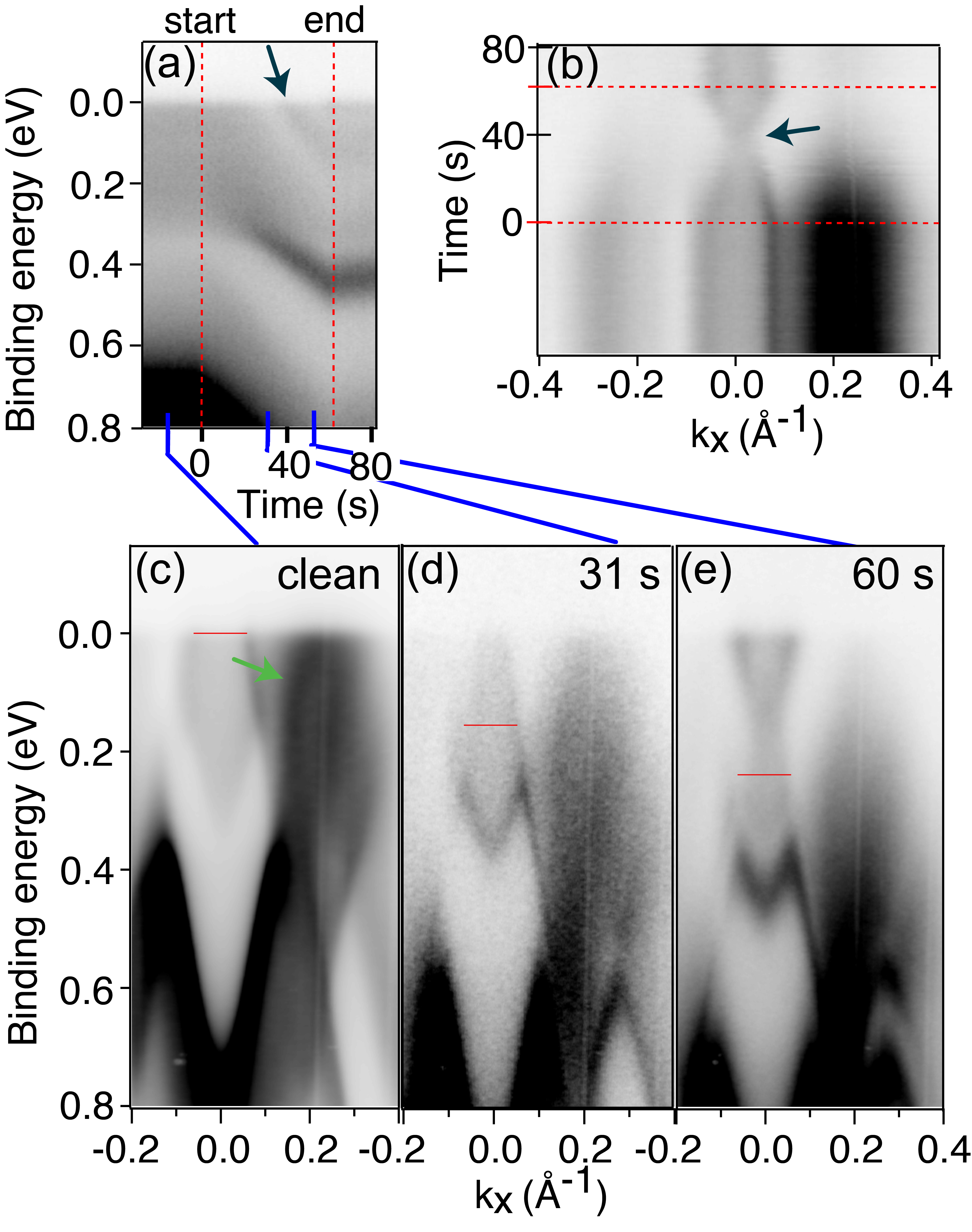}\\
    \caption{Rb dosing at 22 eV. (a) Photoemission intensity in normal emission as a function of binding energy and time while dosing Rb atoms on the surface. Red dashed lines mark the start and end of the Rb dosing. The spectrum is integrated $\pm$ 0.02~\AA$^{-1}$ around normal emission. The black arrow marks the time at which the Dirac point of the topological surface state crosses $E_F$. (b) Photoemission intensity at the Fermi level as a function of time and $k_x$ while dosing Rb atoms onto the surface. The spectrum is integrated $\pm$ 100~meV around the Fermi level. (c)-(e) Photoemission intensity as a function of binding energy and $k_x$ acquired at different moments in time after beginning the Rb dosing. (c) was obtained by integrating the data in (a) from the beginning up until the first blue line and (e) was obtained by integrating  from the last blue marker up until the end of the acquisition. The colour scale of spectra (a) and (c)-(e) was chosen such that some features are saturated.}
    \label{fig:3}
    \end{figure}

Figure \ref{fig:3}(a) shows the photoemission intensity at normal emission for $h\nu=22$~eV as a function of time while dosing Rb atoms onto the surface. The beginning and end of the Rb exposure are marked on the time axis. Figure \ref{fig:3}(b) gives   the corresponding photoemission intensity at the Fermi energy and along $k_x$.   Figures \ref{fig:3}(c)-(e) show snapshots of the electronic structure at selected times. Figure \ref{fig:3}(c) was acquired prior to starting the Rb deposition and is thus very similar to Figure \ref{fig:1}(d). The band forming the VBM along $k_x$ is marked by a green arrow. Upon Rb dosing the observed bands  shift to higher binding energy.  After the end of the exposure, the features gradually move back to lower binding energy. This could either be due to Rb desorption or intercalation. Scanning the UV light spot across the sample shows that the shift takes place over the entire sample surface and is not induced by the UV radiation. The amount of Rb present on the sample surface after 60~s of deposition was estimated to be 0.1 monolayer, determined from the ratio between Rb and Te core level photoemission intensities.     

As discussed above and seen in Figures \ref{fig:3}(a) and (c), the uppermost VB at normal emission appears to be a broad feature between $E_F$ and a binding energy of $\approx$300~meV.  However, when exposing the surface to Rb, the VB features sharpen up and the photoemission intensity is concentrated in a narrow window around the bottom of the band. This can be explained by a quantisation of the VB states due to a downward bending of the bands near the surface, as such a quantization removes the $k_z$ broadening. Naively,  a quantisation upon \textit{downward} bending of the band is expected for the conduction band, not for the VB, but for this class of materials the VB width is so small that the bands become confined once the amount of band bending exceeds the width of the VB, as has been shown for \bise~\cite{Bianchi:2011aa}. The VB quantisation turns the three-dimensional dispersion into two-dimensional sub-bands that no longer disperse in $k_z$. The precise location of the VBM at $\Gamma$ therefore remains inaccessible. We tentatively place the VBM just below the Dirac point. This is consistent with previous observations \cite{Pauly:2012aa} and also with the fact that a weak diffuse photoemission intensity remains inside the lower part of the surface state Dirac cone in Figures \ref{fig:3}(d) and (e).

After a Rb exposure time of 32~s, an additional band can be observed to cross $E_F$  and move to high binding energies in Figure \ref{fig:3}(a) (black arrow). It originates from the Dirac point of the topological surface state. This can be  clearly seen when inspecting the time dependence of the photoemission intensity at $E_F$ shown in Figure \ref{fig:3}(b), where the lines representing the Fermi level crossings of the topological surface states start to converge upon Rb dosing and cross at the same time as the state in Figure \ref{fig:3}(a) becomes visible, bringing the Dirac cone from the hole-doped to the electron-doped regime. Note also that the intensity at $E_F$ around $k \approx \pm$0.3~\AA$^{-1}$ vanishes in Figure \ref{fig:3}(b) upon doping. As these features correspond to the VBM in the $\bar{\Gamma}-\bar{M}$ plane, such a shift suggests that the VBM is at higher binding energy than the Dirac point. Inspecting energy- and $k_x$-dependent spectra at different Rb coverages (Figures \ref{fig:3}(c)-(e)) appears to confirm this.

\begin{figure}
\includegraphics[width=0.45\textwidth]{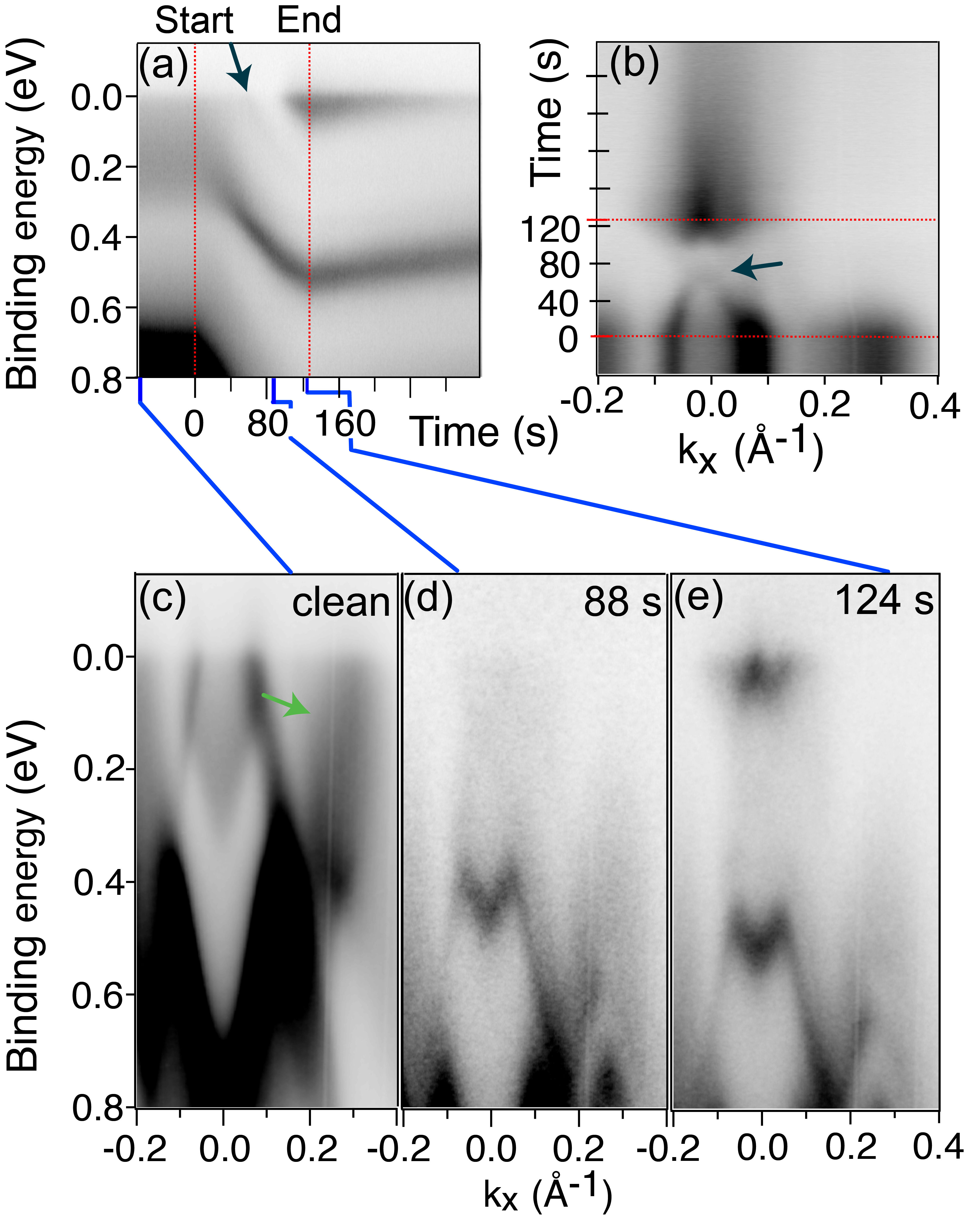}\\
\caption{Rb dosing at 18 eV. The panels correspond to those in Figure \ref{fig:3}. The data for panel (e) has been integrated over 3 s.  }
\label{fig:4}
\end{figure}

A similar behaviour is observed in Figure \ref{fig:4} where doping-dependent ARPES data  acquired at $h\nu=$18~eV is shown. The data are presented in the same way as in Figure \ref{fig:3} and the conclusions about the position of the VBM in the $\bar{\Gamma}-\bar{M}$ plane are largely the same. The band dispersion is less clear, even for the clean surface. Nevertheless, the photoemission intensity from the VBM along $\bar{\Gamma}-\bar{M}$ seen in Figure \ref{fig:3}(b) quickly vanishes in Figure \ref{fig:4}(b) and the band maximum can be clearly placed below the surface state Dirac point. An important difference to Figure \ref{fig:3} is that the Rb exposure was carried out for a longer time (120~s, giving rise to a coverage of 0.3 monolayers). As seen in Figure \ref{fig:3}(e), this leads to an occupation of the conduction band (possibly present as a quantized state). Note that the observation of conduction band states does not permit the determination of the bulk band gap, as the conduction band minimum is not found at $Z$ but close to $\Gamma$  \cite{Nechaev:2015aa}. Furthermore, alkali-induced surface doping does not necessarily conserve the gap size as seen by comparing refs. \cite{Miwa:2015aa} and \cite{Antonija-Grubisic-Cabo:2015aa}, for instance.

Combining the data for the clean and Rb-covered surface, we can conclude the following about the VB in the $\bar{\Gamma}-\bar{M}$ plane: The plane does indeed contain a local VBM. However, its energy is lower than that of the global VBM at the $\Gamma$ point, which we assume to be just below the Dirac point of the surface state. The dispersion of the upper VB in $\bar{\Gamma}-\bar{M}$ shows a maximum 30~meV below $E_F$ for data taken at $h\nu=$22~eV (see Figures \ref{fig:1}(d) and \ref{fig:3}(c)). This feature broadens out upon Rb dosing but we do not find any indication of a higher lying maximum.When determining the position of the surface state Dirac point by extrapolation of the linear dispersion above $E_F$ for the clean surface, we find the $\bar{\Gamma}-\bar{M}$ VBM to be 170~meV $\pm$ 43~meV below the surface state Dirac point. These findings clearly  support the dispersion obtained by GW calculations \cite{Nechaev:2015aa}. We can rule out the DFT results which place the $\bar{\Gamma}-\bar{M}$ VBM above the surface state Dirac point \cite{Hsieh:2009ac,Menshchikova:2011uu}. 

\section{Conclusions}

In conclusion, we have determined the band structure of the topmost VB of \sbte~by synchrotron radiation-based angle-resolved photoemission spectroscopy, with the objective to find the position of the global VBM. We find a predicted VBM in the BZ plane given by the  $\bar{\Gamma}-\bar{M}$ azimuthal direction. The binding energy of this band is below the surface state Dirac point. This is significantly different from predictions by DFT but in line with calculations employing GW corrections. The high binding energy of the local VBM implies that it should be possible to achieve surface state dominated electronic transport near the Dirac point for \sbte. 

\begin{acknowledgments}
This work was supported by VILLUM FONDEN via the Centre of Excellence for Dirac Materials (Grant No. 11744). 
    S. U. acknowledges financial support from VILLUM FONDEN under the Young Investigator Program (grant no. 15375). We thank Anton Tamt\"ogl for fruitful discussions.
\end{acknowledgments}

%\bibliographystyle{apsrev}
%\bibliography{phref}

\end{document}